\RequirePackage{fix-cm}
\documentclass[twocolumn,epjc3]{svjour3}  
\smartqed  % flush right qed marks, e.g. at end of proof
\RequirePackage{graphicx}
\RequirePackage{amsmath}
\RequirePackage{amssymb}
\usepackage{color}
\journalname{Eur. Phys. J. A}
\begin{document}

\title{Target-Normal Single Spin Asymmetries Measured with Positrons
}
%%\subtitle{Do you have a subtitle?\\ If so, write it here}


\author{G.~N.~Grauvogel\thanksref{gw}
    \and
    T.~Kutz\thanksref{gw,mit}
    \and
    A.~Schmidt\thanksref{e1,gw}
    }

\thankstext{e1}{e-mail: axelschmiddt@gwu.edu}


\institute{George Washington University, Washington, DC 20052, USA \label{gw}
\and
Massachusetts Institute of Technology, Cambridge, MA 02139, USA
\label{mit}
}

\date{Received: date / Accepted: date}

\maketitle

\begin{abstract}
Two-photon exchange and the larger class of hadronic box diagrams 
are difficult to calculate without a large degree of model-dependence.
At the same time, these processes are significant radiative corrections in parity-violating electron scattering, in neutron decay, and may even be responsible for the proton's form factor ratio discrepancy. 
New kinds of experimental data are needed to help constrain models and guide future box-diagram calculations. 
The target-normal single spin asymmetry, $A_n$, formed with an unpolarized beam scattering from a target that is polarized normal to the scattering plane, is sensitive to the imaginary part of the two-photon exchange amplitude, and can provide a valuable constraint.
A measurement with both electrons and positrons can reduce sources of experimental error, and distinguish between the effects of two-photon exchange and those of time-reversal symmetry violation. 
This article describes a proposed experiment in Hall A, using the new Super Big-Bite Spectrometer that can cover a momentum transfer range in the critical zone of uncertainty between where hadronic calculations and those based on partonic degrees of freedom are expected to be accurate. 

\keywords{two-photon exchange \and positrons \and single-spin asymmetries \and polarized target}

\end{abstract}

\section{Introduction}

Hadronic box diagrams in elastic electron scattering are difficult to
calculate without significant model dependence. 
Unfortunately, they are also produce significant radiative corrections in a number of measurements, for example, $\gamma Z$-exchange in parity-violating electron scattering and $\gamma W^\pm$-exchange in measurements of beta-decay widths. 
Two-photon exchange (TPE) in elastic electron-proton scattering is hypothesized to be responsible for the discrepancy between unpolarized and polarized extractions of the proton's form factor ratio~\cite{Guichon:2003qm,Blunden:2003sp}. 
All of these applications require a better understanding of box-diagram processes, and new experimental constraints are needed
to help improve theoretical calculations. 
There are several experimental observables that are directly sensitive to box-diagram contributions, and because they provide orthogonal constraints, it is advantageous to pursue a variety.

One such observable is a target-normal single-spin asymmetry (SSA),
denoted here by $A_n$. 
This asymmetry is measured by scattering an unpolarized electron (or positron) beam on a target polarized in
a direction perpendicular to the scattering plane, and comparing
cross sections for ``up'' and ``down'' target polarizations. In the
limit of one-photon exchange, single-spin asymmetries in elastic scattering are forbidden,
so $A_n$ is a direct measure of multi-photon exchange. 

Following the formalism of Ref.~\cite{Carlson:2007sp}, $A_n$ for a 
proton target can be related to the proton's higher-order form factors,
$\delta \tilde{G}_E$, $\delta \tilde{G}_M$, and $\delta \tilde{F}_3$, by
\begin{multline}
A_n = \frac{\sqrt{2\varepsilon(1+\varepsilon)}}{\sqrt{\tau}\left(G_M^2 + \frac{\varepsilon}{\tau}G_E^2\right)} \times
\Bigg[ -G_M \textrm{Im}\left( \tilde{G}_E + \frac{\nu}{m_p^2}\tilde{F}_3 \right)\\
+ G_E \textrm{Im} \left( \tilde{G}_M + \frac{2 \varepsilon \nu}{m_p^2 (1+\varepsilon)} \tilde{F}_3 \right) \Bigg] + \mathcal{O}(\alpha^4),
\label{eq:ffs}
\end{multline}
$G_E$ and $G_M$ are the proton's standard electric and magnetic form factors,
$m_p$ is the mass of the proton, $\tau$ is the dimensionless quantity $Q^2/4M^2$, where $Q^2$ is the magnitude of the squared invariant 4-momentum transfer, $\varepsilon$ is the virtual photon polarization parameter, defined by $\varepsilon^{-1} \equiv 1 + 2(1+\tau)\tan^2\frac{\theta}{2}$, where $\theta$ is the lepton scattering angle in the rest frame of the target, and $\nu$ is the Lorentz-invariant parameter formed by contracting the sum of the initial and final lepton 4-momenta with the sum of the initial and final proton 4-momenta, i.e. $\nu \equiv (p_e + p_{e'})_\mu (p_p + p_{p'})^\mu$. 
Eq.~\ref{eq:ffs} shows that $A_n$ is sensitive to the imaginary parts of the higher-order form factors, meaning that it provides a completely independent constraint from measurements of the unpolarized positron-proton/electron-proton cross section ratio or of polarization transfer, which probe the real parts. 
As with other observables of TPE, the sign of the effect would be reversed in positron scattering.

In addition to two-photon exchange, any process that violates time-reversal symmetry (i.e., T-violating) will also lead to a non-zero $A_n$~\cite{Carlson:2007sp}. T-violating effects would be symmetric for both electrons and positrons~\cite{Christ:1966}. By measuring with both beam species, TPE and T-violation can be distinguished. It is possible for a T-violating electromagnetic current to be generated either by particles with spin 1 or greater \cite{Glaser:1957} (e.g., the deuteron \cite{Schildknecht:1966zz}), or through inelastic scattering~\cite{Christ:1966}. In elastic electron-nucleon scattering, T-violation could only arise from a non-electromagnetic scattering process \cite{Bernstein:1965hj}. The amount of T-violation within the Standard Model would produce an insignificant $A_n$, so a detection would immediately indicate a beyond-Standard Model effect.

There are several different theoretical approaches for calculating hard two-photon
exchange, but they mo\-stly fall into two classes: hadronic (for example, \cite{Tomalak:2014sva,Blunden:2017nby,Ahmed:2020uso} and 
others) and partonic (for example, \cite{Chen:2004tw,Afanasev:2005mp,Kivel:2009eg} and others). 
The hadronic calculations are expected to be most valid at lower momentum transfer, i.e., $Q^2 < 3$~GeV$^2/c^2$,
while the partonic calculations are applicable for very high momentum transfer, i.e., $Q^2 > 5$~GeV$^2/c^2$. While validation of these calculations is needed at all kinematics, measurements in the zone from $3\lessapprox Q^2 \lessapprox 5$~GeV$^2/c^2$, in between the regions of validity for hadronic and partonic calculations, would be especially useful.

\section*{Previous Measurements}

$A_n$ has been measured previously in only a few different reactions, and no measurements have been made in elastic electron-proton scattering with enough precision to resolve a non-zero effect. SLAC published a measurement in 1970 on polarized protons in a butanol target at very forward scattering angles ($\approx 3^\circ$), but could not exclude a zero asymmetry~\cite{Powell:1970qt}. Earlier, less precise results from Frascati, Orsay, and Stanford measured recoil polarization from an unpolarized target, which is an equivalent observable~\cite{DiGiorgio:1965,Bizot:1968,Prepost:1968bow}. A number of experiments have looked for T-violation in inelastic scattering~\cite{Chen:1968mm,Appel:1970ni,Rock:1970sj,Airapetian:2009ab,Katich:2013atq}. Two recent measurements have been made at Jefferson Lab of quasi-elastic scattering from polarized $^3$He \cite{Zhang:2015kna,Long:2019iig}. Precision measurements of elastic scattering on protons have yet to be made with electrons or positrons. 
Ref.~\cite{Zhang:2015kna} measured an asymmetry of a few parts per thousand in $^{3}$He,
which corresponds to an asymmetry of a few percent from polarized neutrons. It would be
reasonable to expect an asymmetry of similar size from polarized protons.

\begin{figure}%[t]
\centering
\includegraphics[width=.9\linewidth]{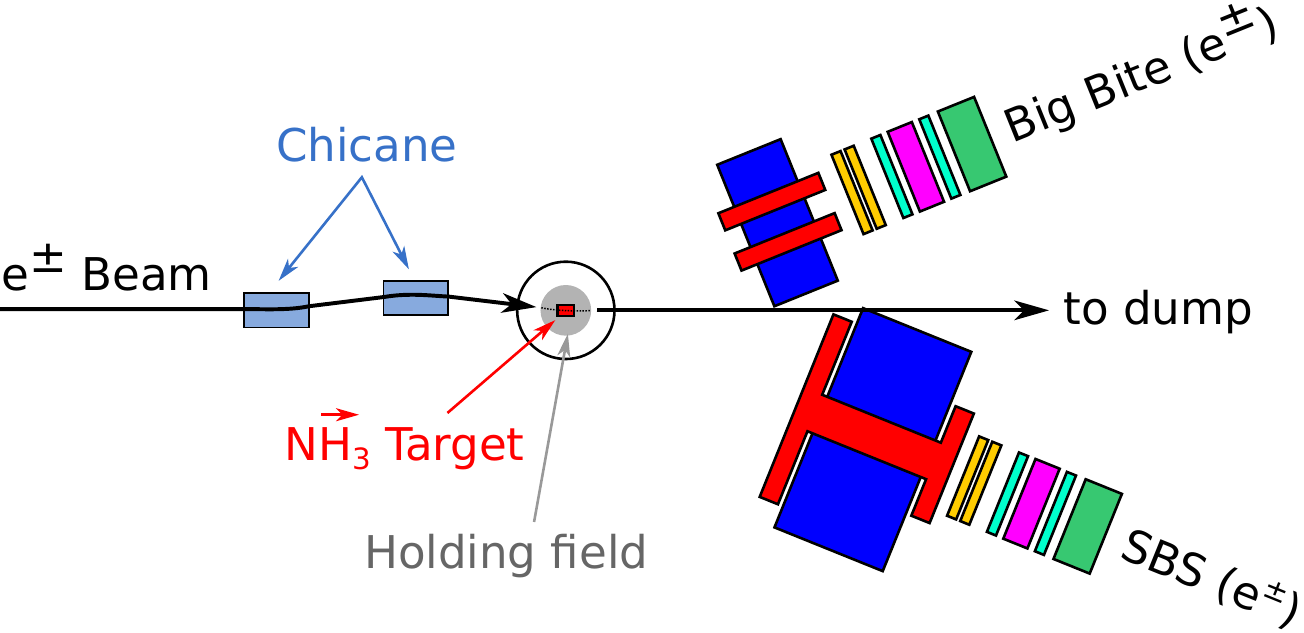}
\caption{Layout of the proposed measurement in Hall A. BigBite and SBS will cover the same scattering angle allowing a simultaneous left/right asymmetry measurement. Due to the bending of charged particles in the holding field, they will sit at slightly different angles in the Hall coordinate system. The holding field will also require that the incoming beam be steered on to the target slightly off axis using a magnetic chicane.
}
\label{fig:layout}
\end{figure}

The measurement of Ref.~\cite{Zhang:2015kna} used the two high-resolution spectrometers in Jefferson Lab's Hall A to simultaneously measure at 17$^\circ$ both left and right of the beam direction. The target was polarized in either the up or the down direction. By using two spectrometers, the target asymmetry was measured as a left/right asymmetry. The target polarization was additionally flipped to reduce systematic uncertainty. This left-right approach has the advantage
of being a simultaneous measurement; both the left and right arms experience the same time-varying beam conditions. As a result, the systematic uncertainty on the raw asymmetry was impressively small, ranging from 0.014--0.026\%. The measurement was statistics limited. For technical reasons, this level of accuracy would be difficult to match with a polarized proton target, but, nevertheless, the Hall A measurement illustrates how this asymmetry is a systematically clean observable.

\begin{table*}
\centering
\caption{\label{tab:proposed} Proposed Measurement Plan}
\begin{tabular}{c c c c c | c c | c c | c c }
\hline \hline
& & & & & \multicolumn{2}{c|}{Big-Bite} &
\multicolumn{2}{c|}{SBS} & & \\
Beam [GeV] & $Q^2$ [GeV$^2$] & $\varepsilon$ & $\theta_e$ [$^\circ$] &
$p_e$ [GeV$/c$] & $\theta_\text{lab}$ [$^\circ$] & $R$ [m] 
& $\theta_\text{lab}$ [$^\circ$] & $R$ [m] &
$\Omega$ [msr] & Days  \\
\hline
6.6 & 4.0 & 0.8696 & 21.2 & 4.34 &
23.2 & 2.1 & 21.2 & 3.0 & 26 & 9 \\
6.6 & 3.0 & 0.9207 & 17.3 & 5.00 &
17.9 & 2.8 & 16.7 & 3.6 & 18 & 3 \\
\hline
4.4 & 3.0 & 0.8065 & 28.6 & 2.80& 
30.3 & 1.6 & 26.9 & 2.5 & 38 & 4 \\
4.4 & 2.0 & 0.9004 & 21.3 & 3.33 &
22.3 & 2.2 & 20.3 & 3.0 & 26 & 1 \\
\hline
2.2 & 2.0 & 0.5600 & 53.2 & 1.13 &
58.9 & 1.1 & 47.5 & 1.6 & 70 & 2 \\
2.2 & 1.0 & 0.8419 & 30.3 & 1.67 &
32.3 & 1.5 & 28.3 & 2.4 & 40 & 0.5 \\
2.2 & 0.5 & 0.9353 & 19.7 & 1.93 &
20.7 & 2.4 & 18.7 & 3.2 & 23 & 0.5 \\
\hline
\textbf{Total} & & & & & & & & & & \textbf{20} \\
\hline
\hline
\end{tabular}
\end{table*}

The measurement of Ref.~\cite{Long:2019iig} also looked at quasi-elastic scattering from $^3$He in Hall A. However, instead of using simultaneous left and right electron arms, the experiment made a coincidence measurement of the  $^3$He$(e,e'n)$ reaction. While this is sensitive to TPE in principle, the requirement of a detected neutron introduces new asymmetry-generating effects, such as final state interactions, meson exchange currents. In fact, Ref.~\cite{Long:2019iig} drew conclusions about the magnitude of non-plane wave reaction effects, rather than TPE or T-violation. It is relevant, however, that Ref.~\cite{Long:2019iig} achieved an approximately $9\%$ \emph{relative} systematic uncertainty, i.e., equivalent to an absolute uncertainty of 0.09\% on 1\% asymmetry. The uncertainty was dominated by the time-variance of the target polarization. We can expect a similar level of systematic uncertainty for a measurements made by a single spectrometer.

\section*{Proposed Measurement}
%%And a reference to a supplement \cref{note:Note1}. And \nameref{note:Note1}.
The addition of a positron source at Jefferson Lab's CEBAF accelerator would make possible a first ever measurement of the single-spin asymmetry $A_n$ in positron scattering. 
We propose here a concept for a measurement of $A_n$ in both electron and positron elastic scattering from polarized protons, which could distinguish TPE and T-violation effects. 
This measurement would also provide the first $A_n$ data on protons for $Q^2>1$~GeV$^2/c^2$, reaching up to 4~GeV$^2/c^2$, in between the regions where hadronic calculations and partonic calculations are expected to be most accurate. 
This is a more ambitious proposal than that first suggested in Ref.~\cite{Schmidt:2017gky}. This is made possible by employing Hall A's new Super BigBite Spectrometer (SBS), paired with the upgraded BigBite Spectrometer (BB), providing larger acceptance than Hall A's HRSs. Both SBS and BB are non-focusing magnetic spectrometers, each with a single dipole magnet bending charged particles upwards from horizontal plane. At their most favorable positions, SBS can have an acceptance over 70~msr~\cite{SBS:CDR} and BB can have an acceptance over 90~msr~\cite{transversity}, compared to the approximately 6~msr acceptance of each HRS~\cite{Alcorn:2004sb}. The use of an even larger acceptance spectr\-ometer---Hall B's CLAS12---is unfeasible due to the need for a strong transverse magnetic field to hold the target polarization. Such a holding field would not be possible inside CLAS12's longitudinal solenoid magnet \cite{Burkert:2020akg}.
A top-view illustration of the experiment layout is shown in Fig.~\ref{fig:layout}.

The unpolarized beam would be directed on a vertically polarized target. Polarized proton targets present a number of technical hurdles, and these are compounded when the polarization is transverse. Maintaining polarization requires a holding field of several tesla in strength; this is much greater than for polarized helium-3 targets, where 30~G = 3 mT is sufficient.
Several different polarized proton targets have been used at Jefferson Lab~\cite{Averett:1999nz,Keith:2003ca,Keith:2012ad,Pierce:2013pua}, with the target from the $g_2^p$ and $G_E^p$ Experiments~\cite{Pierce:2013pua} demonstrating the best performance under high-luminosity conditions. 
In this target, protons are dynamically polarized in frozen beads of ammonia (NH$_3$), within a 2.5--5~T holding field, achieving approximately 70\% average polarization
with a beam current of up to 100~nA. This corresponds to a luminosity of roughly
$10^{35}$~cm$^2$~s$^{-1}$, given a 3~cm long target, and a rough estimate of a 60\% packing fraction for the ammonia beads. 
For this proposal, we assume that a similar target could be produced, which could achieve 66\% polarization at a luminosity of $10^{35}$~cm$^2$~s$^{-1}$. 

The $g_2^p$/$G_E^p$ target was never used with vertical polarization, though there are no technological reasons preventing the target magnet from being oriented vertically. However, the vertical holding field has a major impact on the steering of the beam, and on the trajectories of charged particles emerging from the target. Assuming a total field integral of 1.5~Tm, a 2.2 GeV beam would be deflected horizontally by $\approx 11.4^\circ$ when passing through the target. To correct for this deflection, it would be necessary to use a magnetic chicane to steer the beam onto the target at an angle, so that the target holding field itself could bend the beam back toward the beam dump. Fig.~\ref{fig:layout} shows how this chicane would be oriented in the proposed experiment. One consequence of this design is that the beam will cross the target at a slight angle relative to the hall coordinate system, and this, along with the deflection of outgoing particles must be accounted for when determining the placement of the spectrometers.

\begin{figure}%[t]
\centering
\includegraphics[width=.9\linewidth]{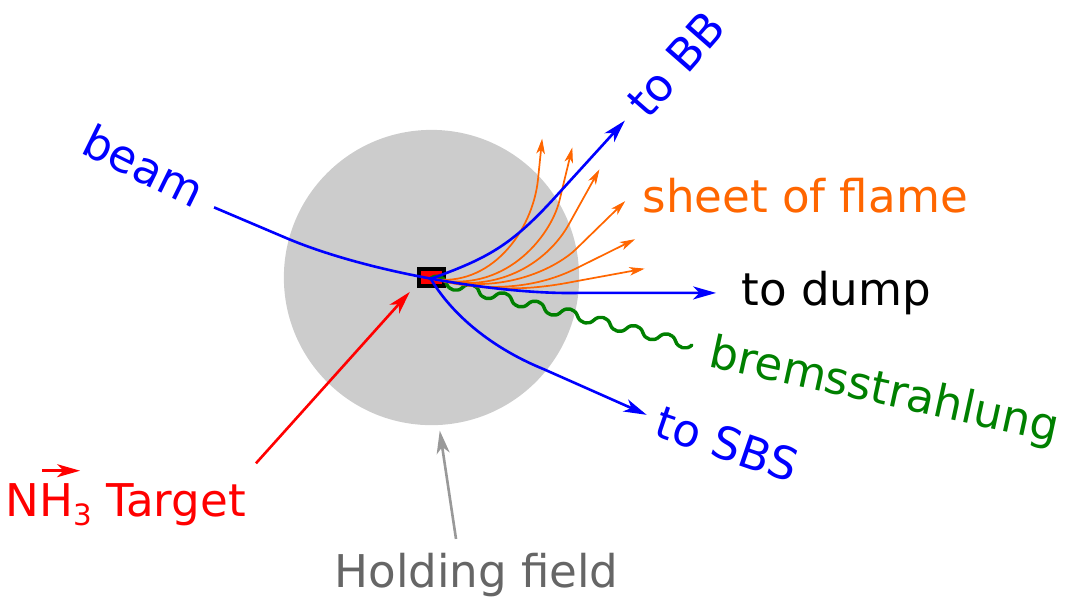}
\caption{An overhead view of the target region showing the curvature of charged particles caused by the holding field. In the proposed measurement, the magnet will be oriented to direct the ``sheet of flame'' to the left side of the beam line, covered by BB. The BB dipole magnet will need to deflect low-momentum sheet-of-flame particles away from the detectors.
}
\label{fig:sheet}
\end{figure}

Transverse polarized targets also have the adverse property of producing a large background of low-energy electrons that is spread in a so-called ``sheet of flame.'' Beam particles that lose energy in the target, e.g., through bremsstrahlung, are deflected strongly in the holding field. For the proposed vertical holding magnet, the sheet of flame will be bent in the horizontal plane, into the aperture of one of the spectrometers, illustrated in Fig.~\ref{fig:sheet}. Fortunately, the sheet-of-flame particles will be much lower momentum than the particles of interest, and the spectrometer dipoles will deflect the sheet out of plane by large angles. Care must be taken to ensure that the sheet of flame misses the active detectors; otherwise they would be swamped with ionization. 

In this experiment, leptons would be scattered elastically from a polarized proton target, and detected in either the SBS or BB spectrometers to the left and right of the beam line, respectively (see Fig.~\ref{fig:layout}). 
The two spectrometers would cover identical scattering angles so that the measurement of $A_n$ could be made from the left-right asymmetry alone, reducing systematic effects from time-varying running conditions. Because the target holding field will deflect the beam and outgoing particles, the placement of the spectrometers in the hall will be at two slightly different angles. We propose orienting the chicane and holding field so that the beam and the sheet of flame are bent to the left, i.e., toward BB and away from SBS, for both electron and positron running modes; to switch between electrons and positrons, the chicane and target holding field polarities will also switch. Conveniently, the target polarization direction can be flipped independently of the holding field polarity. Frequent target polarization flips can be used to further reduce systematics coming from the target and from acceptance differences between the two spectrometers. 

\begin{figure}%[tbhp]
\centering
\includegraphics[width=\linewidth]{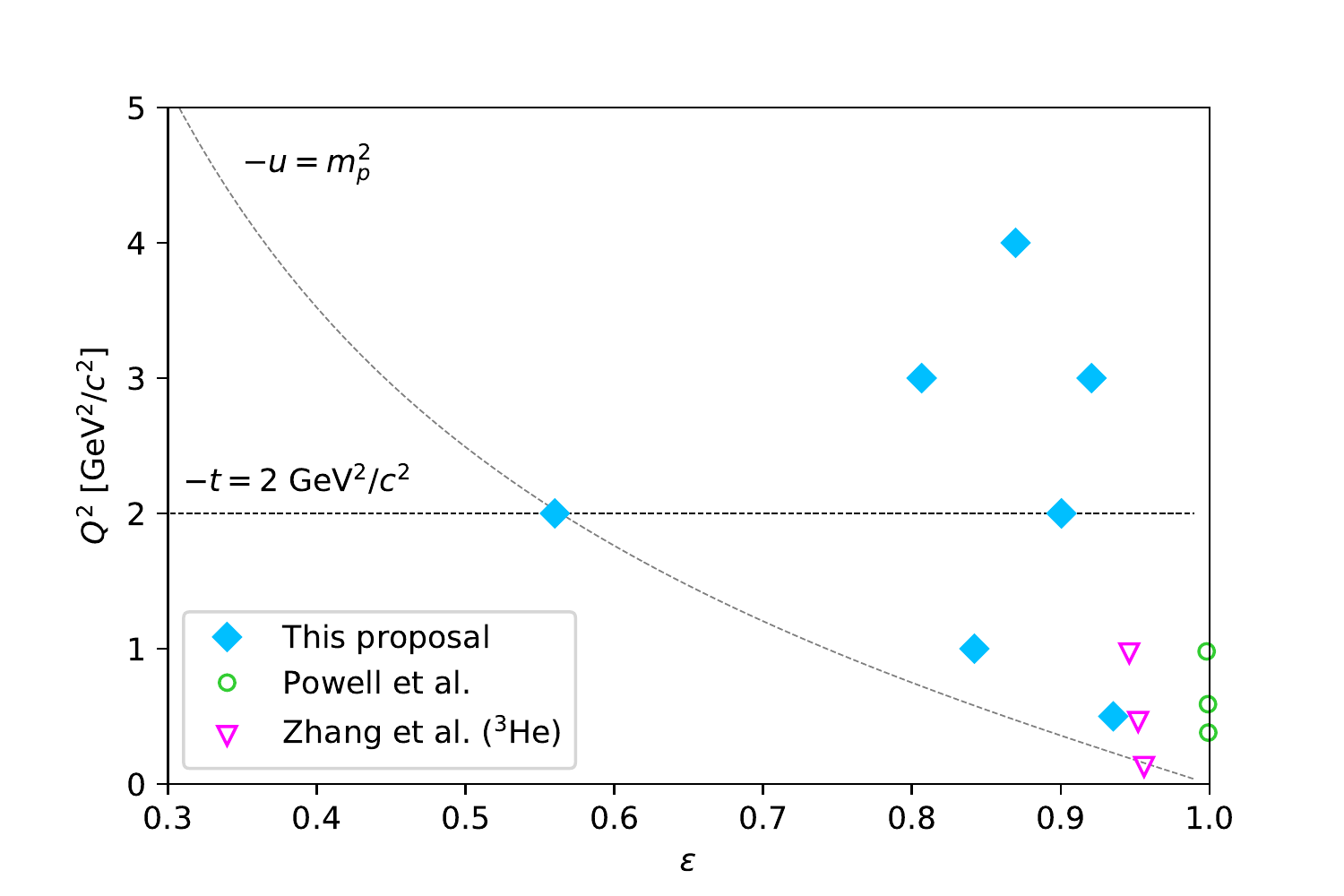}
\caption{
Kinematic map of the proposed measurement (diamonds) in $\varepsilon$,$Q^2$ space. 
The kinematics of the SLAC measurements from 1970 (Powell et al., open circles)~\cite{Powell:1970qt} are shown for comparison. 
The Hall A results for the neutron using a polarized $^3$He target (Zhang et al., open triangles) are from Ref.~\cite{Zhang:2015kna}. 
The partonic calculation of Ref.~\cite{Afanasev:2005mp} is claimed to be valid when Mandelstam variables $|t|>2$~GeV$^2/c^2$ and $|u|>m_p^2$, the region in the upper right area of the map. This region will be probed by three of the proposed kinematic points. }
\label{fig:kin_map}
\end{figure}

SBS and BB would require detector packages optimized for electron detection. Of the currently planned measurements in the Hall A SBS program, the GMn \cite{GMn}, GEn \cite{GEn}, GEn-RP \cite{GEnRP}, and Transversity \cite{transversity} experiments plan to use Big Bite as the electron arm. Though the specifics of their detector packages vary slightly, they all employ GEM planes for tracking, a gas Cherenkov detector for electron ID, a lead glass calorimeter, and a scintillator hodoscope positioned between the pre-shower and shower layers of the calorimeter. We envision using a similar detector configuration here. 

None of the currently planned measurements use SBS for electron detection. To minimize systematics, it would be preferable to make the SBS detector package as similar as possible to that of BB. While suitable GEM tracking stations have been built for SBS, a new gas Cherenkov detector would need to be constructed, as well as a calorimeter. It is possible that the BigCal from the GEp Experiment~\cite{GEP5} could suffice as the calorimeter.

The proposed measurement plan is shown in Table~\ref{tab:proposed}, and consists of a total of 20 running days, with the assumption of 50\% running efficiency built in. Overhead for changing between electron and positron running modes, changing the number of passes in CEBAF, and repositioning the spectrometers is not considered. In this plan, the SBS and BB spectrometers will simultaneously cover $Q^2$ values of 0.5, 1, 2, 3, and 4~GeV$^2/c^2$. The use of three different beam energies, standard first-pass, second-pass, and third-pass CEBAF energies, will allow measurements of $Q^2=2$ and 3~GeV$^2/c^2$ to be made at multiple values of $\varepsilon$. The exact energy per pass is not critical, and this plan could be adjusted slightly depending on how CEBAF is performing when this experiment takes place. The number of days listed includes both electron and positron running, i.e., 4 days means 2 days with an electron beam and 2 days with a positron beam.
To minimize any time-varying systematics, it would be desirable to be able to switch between electrons and positrons frequently, though this would likely need to be balanced against the time needed to flip the polarity of the accelerator.
Similarly, the target polarization direction should be flipped frequently.
Since the target would likely need to be re-annealed every few hours to restore polarization, this would be the sensible timescale for target spin flips. 

The solid angle coverages of SBS and BB depend greatly on how closely the spectrometer magnets can be positioned to the target. 
At forward angles, BB must be moved further away from the target to avoid hitting the downstream beam pipe. 
While the SBS dipole magnet has a cut-out to allow closer placement to the beam pipe, it must still be positioned further away to avoid hitting BB. 
This limits of the angular acceptance of both spectrometers.
Table~\ref{tab:proposed} includes columns for our estimates of the nominal distance to the target for the BB and SBS magnets and the solid angle, $\Omega$, common to both spectrometers based on where they might be placed to avoid hitting each other or the beam pipe. A rigorous optimization of these spectrometer positions would need to be conducted based on the equipment and obstructions in Hall A at the time the experiment is to be run.

As mentioned above, the positioning of the BB detectors must take the sheet of flame into account. 
The BB dipole can produce a field integral of $\approx 1$~Tm, enough to bend the sheet of flame more than 25$^\circ$ out of plane at the most unfavorable forward kinematic point (6.6~GeV beam, $17^\circ$). 
As long as the sheet misses the BB GEMs, it will not pose a problem to the experiment. 
This may require repositioning the GEMs, or reducing their area, in turn decreasing the BB acceptance. 
However, since in the proposed experiment SBS defines the common acceptance, a slight reduction in BB's acceptance may not have any impact on the measurement. In our rate estimates, we assume that BB can safely match the SBS acceptance while avoiding the sheet-of-flame background, but acknowledge that a detailed simulation will be vital to ensure this background can be deflected, blocked, or avoided so that the experiment can be successful.

A kinematic map of the proposed measurement plan in $\varepsilon,Q^2$ space is shown in Fig.~\ref{fig:kin_map}. 

\section*{Statistical Uncertainty}

\begin{figure}%[tbhp]
\centering
\includegraphics[width=\linewidth]{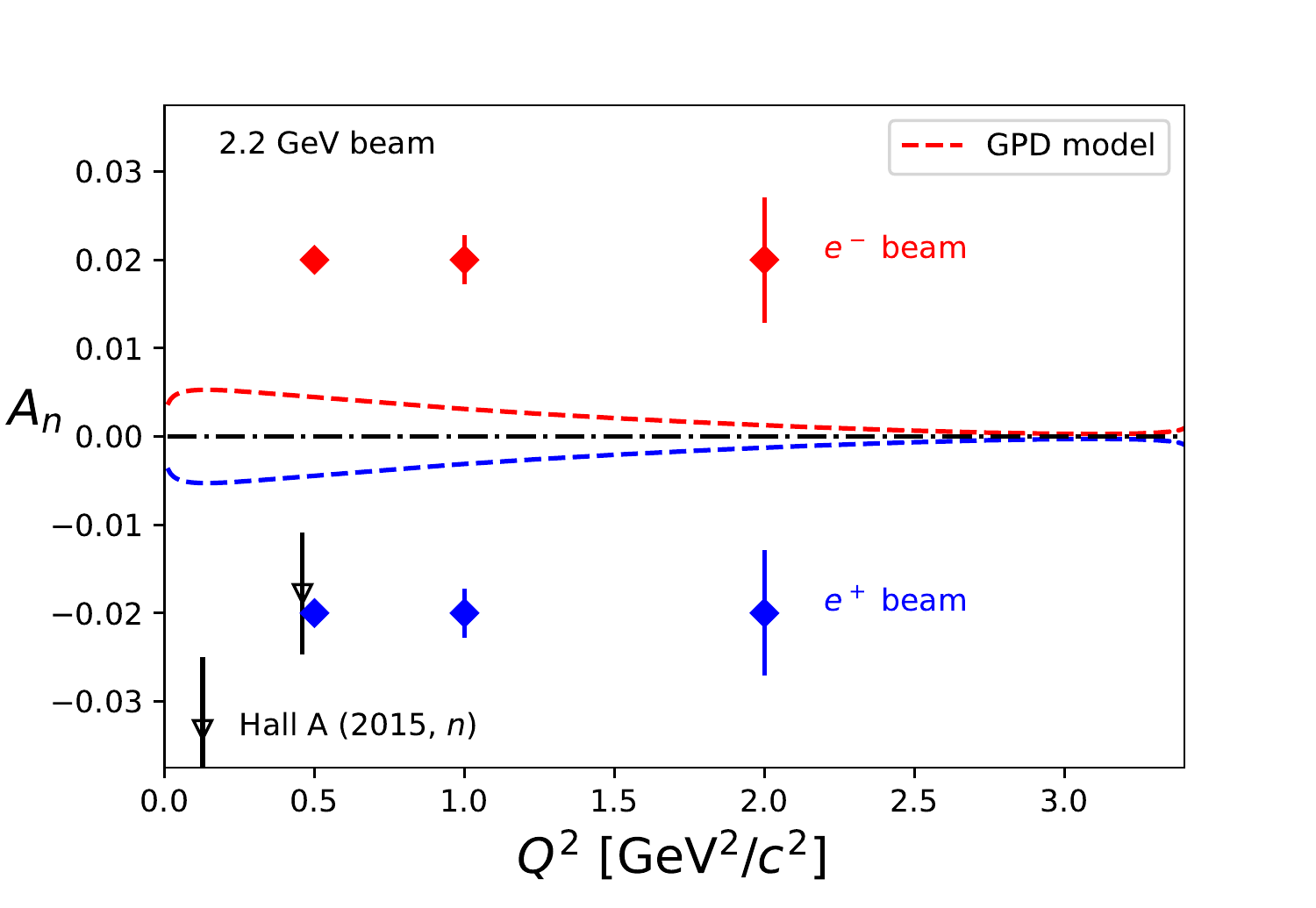}\\
\includegraphics[width=\linewidth]{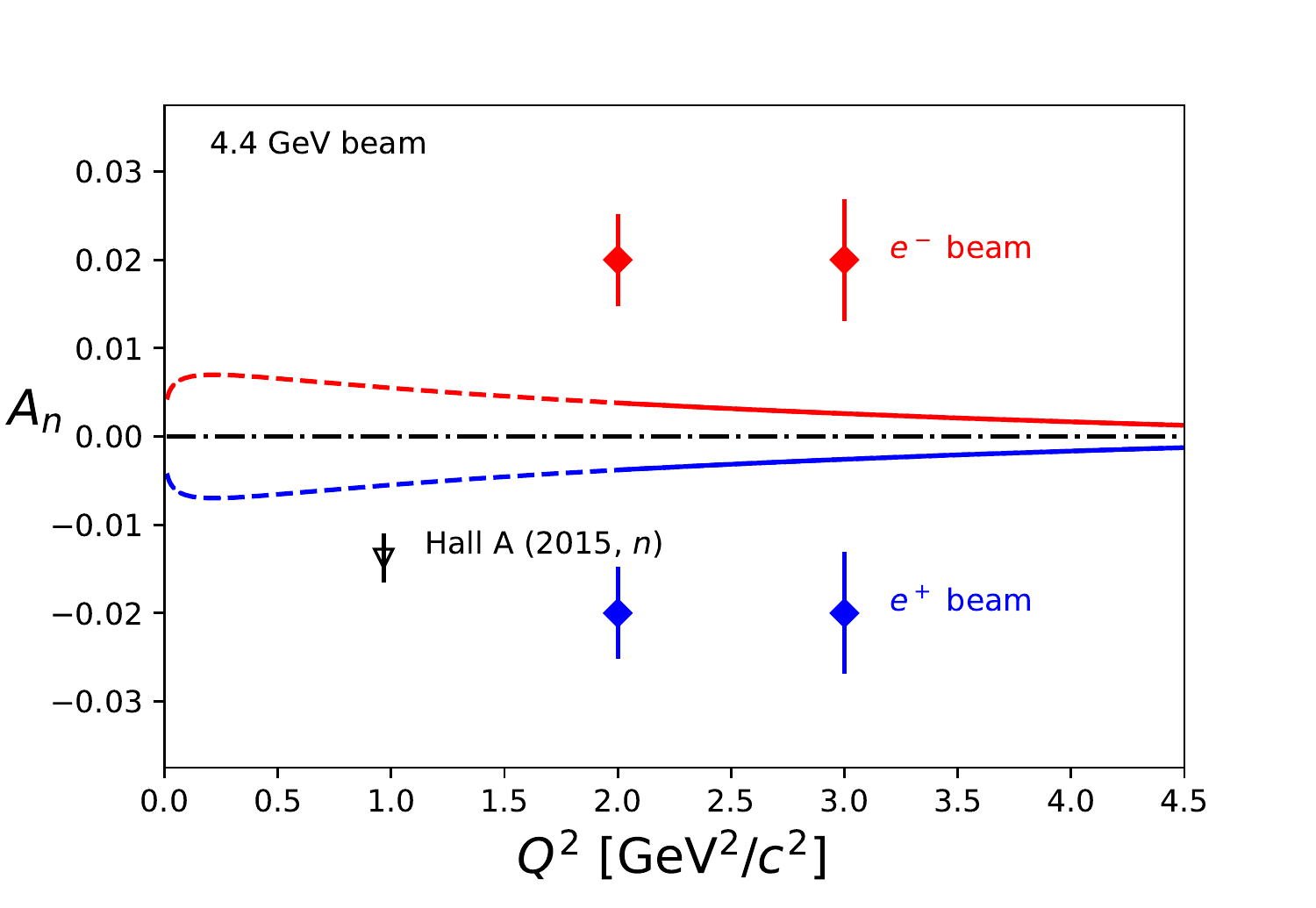}\\
\includegraphics[width=\linewidth]{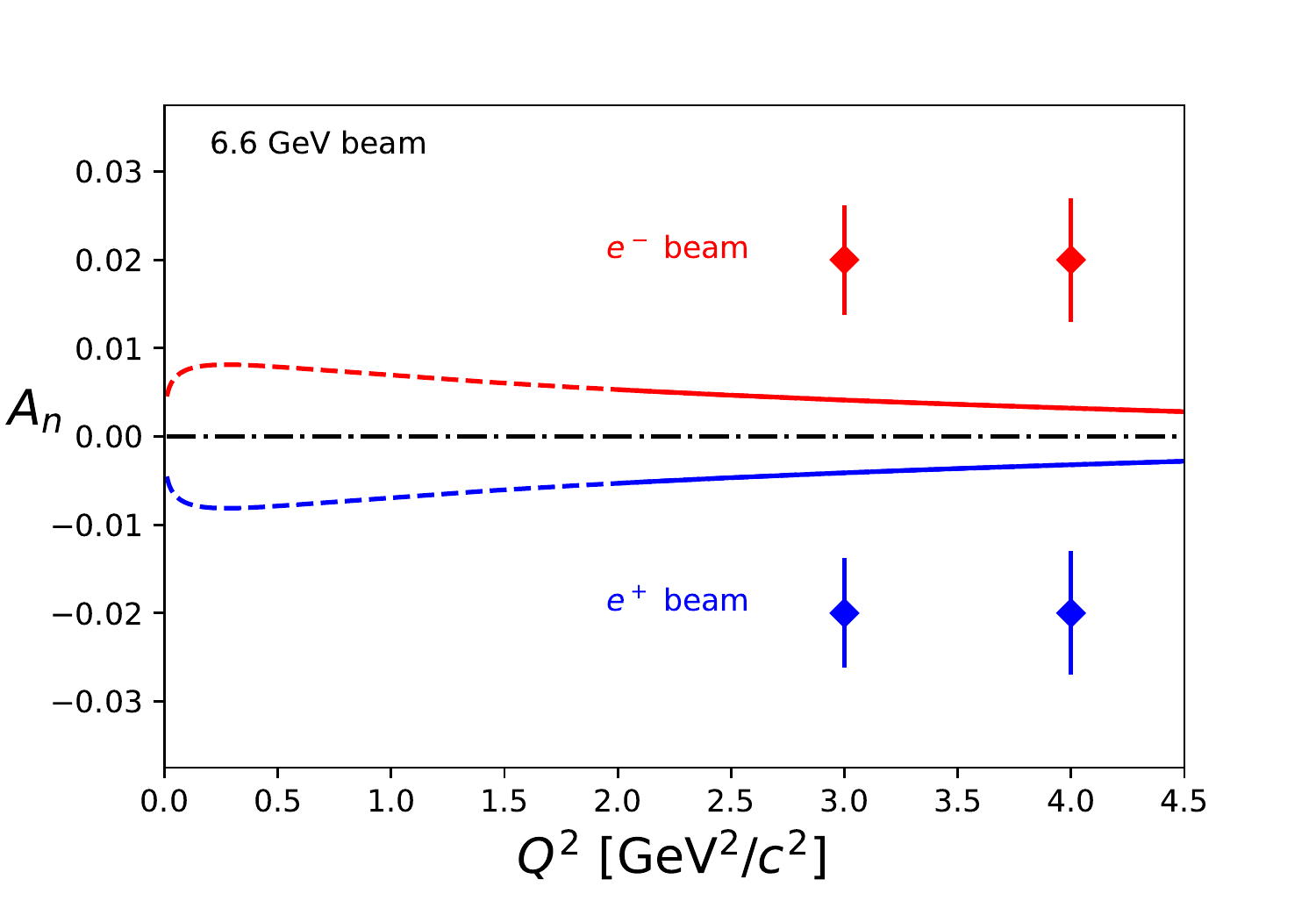}
\caption{
Anticipated statistical precision of the proposed measurement for the three different beam energies. The projected statistical uncertainties are approximately 0.75\% or better for all kinematic points. The Hall A results for the neutron using a polarized $^3$He target are from Ref.~\cite{Zhang:2015kna} (open triangles). Slightly different beam energies were used, and we place the points on the panel with the closest beam energy.
The curves show the predictions of the Gaussian GPD model of Afanasev et al.~\cite{Afanasev:2005mp} for both an electron (red) and positron (blue) beam. The calculation is claimed to be valid for $-t>2$~GeV$^2$ and $-u>m_p^2$. When those conditions are not met, we show the calculation as a dashed line rather than a solid line.}
\label{fig:reach}
\end{figure}

The statistical precision for the proposed measurement was estimated as:
\begin{equation}
  \delta A_n = \frac{1}{PD\sqrt{2\frac{d \sigma}{d\Omega} \Omega \mathcal{L} T \epsilon}},
\label{eq:prec}
\end{equation}
where $P$ is the target polarization (assumed 66.7\%), $D$ is the target dilution factor (assumed 50\% for NH$_3$ based on the experience of Ref.~\cite{Jones:2006kf}),
$\frac{d \sigma}{d\Omega}$ is the elastic scattering cross section, $\Omega$ is the single spectrometer
acceptance, $\mathcal{L}$ is the luminosity ($10^{35}$~cm$^{-2}$s$^{-1}$), $T$ is the run time per
target polarization setting, and $\epsilon$ is the running efficiency (assumed 50\%). For these estimates, SBS and BB were assumed to cover the common angular acceptance listed in Table~\ref{tab:proposed}.
The proposed measurement would have approximately 0.75\% statistical precision or smaller on $A_n$. The results of Ref.~\cite{Zhang:2015kna} indicate that, at least for the neutron, $A_n$ is a percent-level asymmetry. 
Fig.~\ref{fig:reach} shows the size of the statistical error bars for the three different beam energies in comparison to theoretical predictions as well as the neutron results from Ref.~\cite{Zhang:2015kna}. The GPD-based partonic calculation of Ref.~\cite{Afanasev:2005mp} predicts a sub-percent asymmetry in the proposed kinematics, which would be difficult to discern for the highest $Q^2$ points. However, we note that the validity of partonic calculations in the kinematics is an open question, that $A_n$ for the neutron was measured to be several percent, and that the proposed measurement would have higher precision and much greater kinematic reach. 

\section*{Systematics}

The proposed measurement would have several sources of systematic uncertainty to overcome.
The dominant source would be the time-variation of the target polarization. 
The target polarization is one of the multiplicative factors that is needed to extract $A_n$ from the measured count-rate asymmetry, and uncertainty in this polarization goes directly into uncertainty in $A_n$. 
For the target described in Ref.~\cite{Pierce:2013pua}, polarization was monitored continuously through NMR, and a similar procedure would be vital for the proposed measurement. 
The stability of the NMR system would be critical in order to ascertain the charge-weighted target polarization for every measurement setting. 
Ref.~\cite{Pierce:2013pua} observed that in between annealings, the target polarization degraded significantly with accumulated dose: though peak polarizations of 90\% were achieved, this would degrade steadily, and the average polarization obtained was only 70\%. 
The $g_2^p$-Experiment claimed an uncertainty of 2--4.5\% on the target polarization~\cite{Zielinski:2017gwp}. For the proposed measurement, this would translate to a \emph{relative} uncertainty on the asymmetry, i.e. a 5\% target polarization uncertainty would produce an uncertainty of 0.05\% on a 1\% asymmetry. This is comparable to what was achieved with the polarized $^3$He target in Ref.~\cite{Long:2019iig}.
Note that the time-dependence would be somewhat mitigated, since we propose a \emph{simultaneous} left/right measurement. 

Another systematic uncertainty would come from the knowledge of the dilution factor of the NH$_3$ target. Elastic scattering events from polarized protons in the target would be accompanied by a quasi-elastic background from nucleons in nitrogen nuclei. This background can be subtracted since the signal is strongly peaked at $W^2 = m_p^2$, where $W$ invariant mass of the final state hadronic system, while the background is broad and smoothly varying due to Fermi motion. However, there would be systematic uncertainty associated with this subtraction, and it would be common to both positron and electron measurements, i.e., it would have no effect on T-violation, but would bias the measurement of TPE. The potential bias of this dilution effect can be reduced by  enriching the target material with $^{15}$N, which is only one proton-hole shy of doubly-magic $^{16}$O. 
By contrast $^{14}$N has both an unpaired proton and an unpaired neutron. A $^{15}$N-enriched target and a subtraction of the quasi-elastic background was employed in a measurement in Hall C~\cite{Jones:2006kf} that achieved a 1\% uncertainty on the dilution factor. While the success of this approach depends on the momentum resolution of the spectrometer used, one could expect similar uncertainty for the experiment proposed here.

Other systematics, including differences in spectrometer efficiencies, spectrometer alignment, and beam current monitoring, would be mitigated by making a simultaneous left/right asymmetry, and by flipping the target polarization.

\section*{Summary}

The target-normal single spin asymmetry $A_n$ is sensitive to the imaginary part of the two-photon exchange amplitude, and has never been measured on a polarized proton target. 
By using both a positron and an electron beam, not only can a measurement of $A_n$ provide new independent constraints on TPE, but 
some experimental systematics can be reduced, and the effects of TPE can be distinguished from those of T-violation. 
Using the new SBS in Hall A in conjunction with the refurbished BB spectrometer, a 20-day measurement would allow a comprehensive 
scan of $Q^2$ at modest $\varepsilon$, and push up to $Q^2=4$~GeV$^2/c^2$, in-between the region of validity for hadronic calculations and those using GPDs or other partonic approaches. These data would constrain models of TPE, and assist in the effort to understand and model electroweak hadronic box processes.

\iffalse
\label{intro}
Your text comes here. Separate text sections with
\section{Section title}
\label{sec:1}
Text with citations \cite{RefB} and \cite{RefJ}.
\subsection{Subsection title}
\label{sec:2}
as required. Don't forget to give each section
and subsection a unique label (see Sect.~\ref{sec:1}).
\paragraph{Paragraph headings} Use paragraph headings as needed.
\begin{equation}
a^2+b^2=c^2
\end{equation}

\begin{figure}
  \includegraphics{example.eps}
\caption{Please write your figure caption here}
\label{fig:1}       % Give a unique label
\end{figure}
\begin{figure*}
  \includegraphics[width=0.75\textwidth]{example.eps}
\caption{Please write your figure caption here}
\label{fig:2}       % Give a unique label
\end{figure*}
\begin{table}
\caption{Please write your table caption here}
\label{tab:1}       % Give a unique label
\begin{tabular}{lll}
\hline\noalign{\smallskip}
first & second & third  \\
\noalign{\smallskip}\hline\noalign{\smallskip}
number & number & number \\
number & number & number \\
\noalign{\smallskip}\hline
\end{tabular}
\end{table}

\fi

\begin{acknowledgements}
We thank J.~D.~Maxwell of the Jefferson Lab target group for helpful discussions about working with transversely polarized targets.
\end{acknowledgements}

\bibliographystyle{spphys}       % APS-like style for physics
\bibliography{references}   % name your BibTeX data base

%\begin{thebibliography}{}
% etc
%\end{thebibliography}

\end{document}